\documentclass[sigconf]{acmart}

\AtBeginDocument{%
  \providecommand\BibTeX{{%
    \normalfont B\kern-0.5em{\scshape i\kern-0.25em b}\kern-0.8em\TeX}}}

\setcopyright{acmcopyright}
\copyrightyear{2023}
\acmYear{2023}
\acmDOI{XXXXXXX.XXXXXXX}

\acmConference[DLG-KDD'23]{The Tenth International Workshop on Deep Learning on Graphs }{August 07, 2023}{Long Beach, CA}
\acmPrice{15.00}
\acmISBN{978-1-4503-XXXX-X/18/06}

\usepackage{bmpsize}
\usepackage{microtype}
\usepackage{graphicx}
\usepackage{subfigure}
\usepackage{booktabs} 
\usepackage{adjustbox}
\usepackage{xspace}
\newcommand{\model}{OCTAL\xspace}
\newcommand{\RN}[1]{%
  \textup{\uppercase\expandafter{\romannumeral#1}}%
}
\usepackage{multirow, makecell}

\usepackage{tikz}
\usepackage{tikzscale}
\usepackage[compat=1.1.0]{tikz-feynman}
\usetikzlibrary{shapes,decorations.pathmorphing}
\tikzset{node distance=0.5cm, every edge/.style={draw,
->,thick},snake it/.style={decorate, decoration=snake}}
\tikzstyle{player1}=[circle,draw=black!100,fill=white!20,thick,inner sep=0pt,minimum size=8mm]
\tikzstyle{player3}=[circle,draw=black!100,fill=none,thick,inner sep=0pt,minimum size=10mm]
\tikzstyle{safeReach}=[circle,draw=white!100,fill=white!20,thick,inner sep=0pt,minimum size=8mm]
\tikzstyle{player2}=[rectangle,draw=black!100,fill=white!20,thick,inner sep=0pt,minimum size=7mm]
\tikzstyle{otherLabel}=[circle,draw=black!100,fill=white!20,thick,inner sep=0pt,minimum size=20mm]
\tikzstyle{cpreLabel}=[rectangle,draw=black!100,fill=white!20,thick,inner sep=0pt,minimum size=40mm]

\usepackage[utf8]{inputenc}
\usepackage[T1]{fontenc}   
\usepackage{hyperref}
\usepackage{url}
\usepackage{booktabs}
\usepackage{amsfonts}
\usepackage{nicefrac}
\usepackage{microtype}
\usepackage{xcolor} 
\usepackage{subfiles}
\usepackage{amsmath}
\usepackage{mathtools}
\usepackage{multirow}
\newcommand{\proj}{OCTAL }

\begin{document}

\title{OCTAL: Graph Representation Learning for LTL Model Checking}



\author{Prasita Mukherjee, Haoteng Yin}
\affiliation{%
  \institution{Department of Computer Science, Purdue University}
  \city{West Lafayette}
  \country{USA}
}
\email{{mukher39,yinht}@purdue.edu}


\begin{abstract}
  Model Checking is widely applied in verifying the correctness of complex and concurrent systems against a specification. Pure symbolic approaches while popular, suffer from the state space explosion problem due to cross product operations required that make them prohibitively expensive for large-scale systems and/or specifications. In this paper, we propose to use graph representation learning (GRL) for solving linear temporal logic (LTL) model checking, where the system and the specification are expressed by a B{\"u}chi automaton and an LTL formula, respectively. A novel GRL-based framework \model, is designed to learn the representation of the graph-structured system and specification, which reduces the model checking problem to binary classification. Empirical experiments on two model checking scenarios show that \model achieves promising accuracy, with up to $11\times$ overall speedup against canonical SOTA model checkers and $31\times$ for satisfiability checking alone.
\end{abstract}


\maketitle 

\section{Introduction \label{sec:Introduction}}
\begin{figure*}

\subfigure[$B$ accepting $(a \: U \: !b)$.\label{fig:BA}]{
\begin{minipage}{0.13\textwidth}
\centering    
\includegraphics[height=2cm]{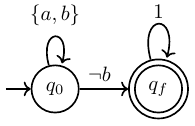}
\end{minipage}
}
\hspace{6mm}
\subfigure[Bipartite graph $\mathcal{G}$ of $B$.\label{fig:BP}]{
\begin{minipage}{0.14\textwidth}
\centering
\includegraphics[height=2.1cm]{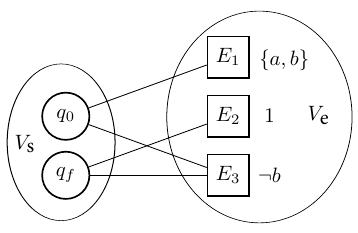}
\end{minipage}
}
\hfill
\subfigure[Expression Tree $\mathcal{T}$ of $\phi: (a \: U \: !b)$.\label{fig:Tree}]{
\begin{minipage}{0.1\textwidth}
\centering
\includegraphics[height=2cm]{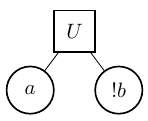}
\end{minipage}
}
\hspace{8mm}
\subfigure[$\mathcal{C} = \mathcal{G} \cup \mathcal{T}$\label{fig:Uni}]{
\begin{minipage}{0.12\textwidth}
\centering
\includegraphics[height=2cm]{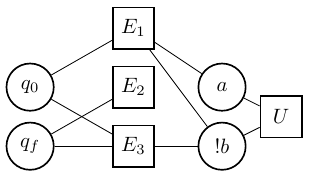}
\end{minipage}
}
\hfill
\subfigure[$B' = B \times B_{\neg \phi}$\label{FigurelabelBACrossProd}]{
\begin{minipage}{0.23\textwidth}
\centering
\includegraphics[height=3.3cm]{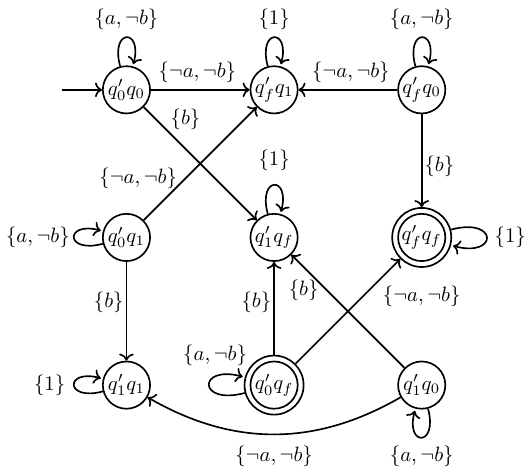}
\end{minipage}
}
\vspace{-5mm}
\caption{Illustration of system $B$, bipartite format of $B$ in graph $\mathcal{G}$, expression tree $\mathcal{T}$ for specification $\phi$,  the unified framework $\mathcal{C}$ approximating the cross product  $B' = B \times B_{\neg \phi}$ with polynomial complexity of $|B+\phi|$, and $B'$ as computed by traditional MCs. \label{fig:overall}}
\vspace{-3mm}
\end{figure*}

Model checking \cite{clarke2018model} is defined as the problem of deciding whether a specification holds for all executions of a system. Generally, formal specifications are expressed using temporal logic formulae like LTL \cite{LTLPnueli}, CTL \cite{LTLPnueli}, etc. The system/model is expressed using automata like B{\"u}chi \cite{Bchi1990}, Muller, Kripke structures \cite{kripkeStructure}, or, Petri nets \cite{petriNet} to express concurrent systems. Given the system $B$ and specification $\phi$, model checking can automatically verify whether the system satisfies the specification by computing automaton $B_{\neg\phi}$, followed by the cross product of $B$ and $B_{\neg\phi}$, and then checks for the emptiness of the product (refer to Figure \ref{FigurelabelBACrossProd}). However, this approach suffers from the state space explosion problem \cite{ValmariExplosion}, which severely hinders the performance of a model checker. Methods such as partial order reduction \cite{partialOrderMC}, symmetry \cite{symmetryLTL}, bounded model checking \cite{Biere99symbolicmodel}, have been proposed to address this problem, but it remains hard in general and constitutes a major bottleneck in deploying model checking for real-world applications.

Recently, machine learning (ML) methods \cite{BehjatiRI,ZhuMC} have gained success in symbolic model checking, giving results in cases where traditional model checkers (MCs) time out. This makes them useful in scenarios where traditional MCs time out/fail to solve, with \emph{speed} and \emph{efficiency} being the key factors. For instance, in software verification, traditional MCs are not always a viable choice due to their high computational cost. Consider a large software development effort, where it would be favorable to use MC to ascertain a higher degree of correctness than pure testing. In such large-scale deployments, classical MCs often take a prohibitively long time for verification, especially when the system/specification has an exponential state space. In this case, only ML-based MCs can provide a practical solution, which broadens the applicability of MCs by trading off some amount of accuracy guarantees for better running time and scalability, which is particularly promising for large systems and/or specifications. 

In this work, we address Model Checking through representation learning. Due to the structural essence of the input, model checking can be naturally formulated into graph tasks. This motivates us to propose a novel graph representation learning (GRL) based framework, \model, to tackle this challenging problem. In \model, the system is expressed as a B{\"u}chi automaton $B$ (Figure \ref{fig:BA}) and the specification with an LTL formula $\phi$ (Figure \ref{fig:Tree}). Then, \model determines whether $B$ satisfies $\phi$ by reducing the problem to binary classification on the graph union of $B$ and $\phi$ (Figure \ref{fig:Uni}).

We performed extensive experiments on \model, traditional MCs, and neural network baselines for two scenarios of LTL model checking, in terms of both accuracy and speed on four datasets: two constructed from open competition RERS19 \cite{rers2019} and two others specifically constructed for this project. Experimental results show that \model consistently achieves $\sim$90\% precision, recall, and accuracy indicating its generalization ability on unseen data, on varied length specifications, and its high utility in practice. In general, \model is up to $11\times$ faster than the state-of-the-art (SOTA) traditional MCs, and achieves at least $31\times$ speedup in terms of satisfiability checking alone. Our major contributions can be summarized as follows: 1) LTL model checking is firstly formulated as a representation learning task, where $B$ and $\phi$ are expressed as graph-structured data. 2) Four datasets are constructed for LTL model checking benchmark: \texttt{SynthGen}, \texttt{RERSGen} correspond to the traditional model checking scenario, and \texttt{SynthSpec}, \texttt{RERSSpec} correspond to the special model checking scenario.

\vspace{-3mm}

\section{\model: \texorpdfstring{LTL M\underline{o}del \underline{C}hecking via Graph Represen\underline{ta}tion \underline{L}e}earning} 
\label{sec:Method}

\model determines whether a system $B$ (B{\"u}chi automaton) satisfies a specification $\phi$ (LTL formula) through their unified graph representation, which is summarized in Figure \ref{fig:overall}. We formulate the problem as supervised learning on graphs, where the inputs $B$, $\phi$ and  label (`0/1') are provided during training. Here, `1' indicates that $B$ satisfies $\phi$ and `0' otherwise. Appendix \ref{appx:symb} provides detailed explanation on B{\"u}chi automaton and traditional LTL Model Checking with concrete examples.

\begin{table}
\centering
\caption{Operands/Variables in specification $\phi$ and system $B$.}
\label{tab:opLogic}
\vspace{-2mm}
\centering
\resizebox{0.99\columnwidth}{!}{
  \begin{tabular}{l|cccc|ccc}
    \toprule
\multirowcell{2}[-5pt][l]{\textbf{Operands/}\\\textbf{Variables}} & \multicolumn{4}{c|}{Specification $\phi$} & \multicolumn{3}{c}{System $B$} \\ 
\cmidrule(r{0.5em}){2-8}
& \textbf{$\mathcal{A}$} & \textbf{$\mathcal{\tilde A}$} & \textbf{true($1$)} &  \textbf{false($N$)} & \textbf{$\mathcal{A}$} & \textbf{$\mathcal{\neg A}$} & \textbf{true($1$)} \\
\midrule
\textbf{Meaning} & $a$ to $z$ & $!a$ to $!z$ & $a \: | \: !a$, $a \in \mathcal{A}$ & $a \: \& \: !a$, $a \in \mathcal{A}$ & $a$ to $z$ & $\neg a$ to $\neg z$ & $a \: | \: \neg a$, $a \in \mathcal{A}$ \\
\midrule
\textbf{Cardinality} & 26 & 26 & 1 & 1 & 26 & 26 & 1\\
\bottomrule 
\end{tabular}}
\vspace{-3mm}
\end{table}

\subsection{Variables and Operators}
The systems and specifications we deal with are constructed from the operands/variables $\mathcal{A}$, operators $\mathcal{O} =\{G, F, R, W, M, X, U, !, \&, |\}$ and special variables true($1$) and false($N$), the specifics of which are described in Table \ref{tab:opLogic} above and Table \ref{tab:symbol} in Appendix \ref{appx:symb}. Each variable and operator has a distinct meaning and share across $B$ and $\phi$. A variable has its true or negated form (noted as $\mathcal{\tilde A}$ or $\mathcal{\neg A}$).

\subsection{Representation of System and Specification}

\paragraph{System Graph $\mathcal{G}$}
We represent $B$ as a bipartite graph $\mathcal{G} = (V_{\mathcal{G}},E_{\mathcal{G}})$, where $V_{\mathcal{G}} = V_{s} \cup V_{e}$ and $E_{\mathcal{G}} \subseteq V_{s} \times V_{e}$. Here, $V_{s}$ are the states of $B$, and $V_{e}$ are the transitions of $B$. There is an edge between $v_i \in V_{s}$ and $v_j \in V_{e}$ if and only if $v_i$ is the source or destination state of the transition $v_j$ in $B$. $\mathcal{G}$ is undirected and the nature of $v_{i}$ being a source or destination vertex is captured in its node encoding.

Figures \ref{fig:BA} and \ref{fig:BP} illustrate $B$ and the corresponding bipartite graph $\mathcal{G}$. The two states $q_{0}$ and $q_{f}$ form the set $V_{s}$, and the transitions $E_{1}, E_{2}$ and $E_{3}$ form the set $V_{e}$. Since $q_{0}$ is a source and destination state for $E_{1}$, and a source state for $E_{3}$, there is an edge between $q_{0}$ and $E_{1}$, and $q_{0}$ and $E_{3}$ respectively. This is analogously followed by the rest of the graph. The intuition behind representing $B$ as a bipartite graph is to capture the transition labels. Since we aim to learn the overall representation of a given system, both states and transitions in $B$ play an essential role here. A state transitions into another state if and only if the transition label is satisfied. To learn the semantics of transitions and their corresponding labels, we map transitions as nodes as shown in Figure \ref{fig:BP} accordingly, and therefore can obtain the representation for them, which is a function of the labels pertaining to the transition.

\paragraph{Specification Graph $\mathcal{T}$}
Every LTL formula can be represented as an expression tree $\mathcal{T} = (V_{\mathcal{T}},E_{\mathcal{T}})$ (see Figure \ref{fig:Tree}), which is constructed based on the precedence and associativity of the operators in LTL formulae, described as follows: 1) $\phi$ is converted to its postfix form, which is used to construct the expression tree; 2) The operators exhibit right associativity, where the unary operators $\{!, G, F, X\}$ have the highest priority. 3) The binary temporal operators $\{U, R, W, M \}$ have the second highest priority, and the boolean connectives $\{\&, |\}$ have the lowest priority. 

$V_{\mathcal{T}}$ constitutes the operators and operands of $\phi$, and $E_{\mathcal{T}} \subseteq V_{\mathcal{T}} \times V_{\mathcal{T}}$. $\phi$ is represented in Negation Normal Form (NNF) \cite{nnf}, which would place $!$ \emph{only} before the operands. This allows us to represent $!a$ as a variable in $\tilde{A}$ and eliminate the $!$ operator. For example, the NNF equivalent of $!(a \: U \:b)$ is $!a \: R \: !b$. Here, the $!$ operator is present only before $a$ and $b$ in the NNF equivalent of the formula $!(a \: U \: b)$. Another compelling reason for representing $\phi$ in NNF is that transitions of $B$ comprise of labels in true($1$), $\mathcal{A}$, and $\mathcal{\neg A}$. Representing negation \emph{only} before variables and eliminating the $!$ operator from $\phi$ enables the shared representation for variables across $B$ and $\phi$. As a result, there is no semantic difference between $\mathcal{\neg A}$ and $\mathcal{\tilde A}$: $\neg a \in \mathcal{\neg A}$ may occur in a transition label of $B$ and $!a \in \mathcal{\tilde A}$ may occur in a leaf node of $\phi$, but both $\neg a$ and $!a$ signify that $a$ does not hold.  
\vspace{-3mm}
\subsection{Bridging System and Specification via Graph Union}
To establish the relation between graphs of system $\mathcal{G}$ and specification $\mathcal{T}$, we propose to construct the unified graph $\mathcal{C}$ to model them jointly. Traditional approaches of model checking computes the intersection of $B$ and $B_{\neg \phi}$ by their cross product, which results in an automaton $B'$ whose states are the product of the states of $B$ and $B_{\neg \phi}$ (Figure \ref{FigurelabelBACrossProd}), and its transitions depend on the transition labels of both $B$ and $B_{\neg \phi}$. Since our main goal is to avoid constructing $B_{\neg \phi}$ and thus $B'$, we directly feed the input graph without products to \proj and aim to use neural networks to learn the latent correspondence by combining $\mathcal{G}$ and $\mathcal{T}$ as a joint framework $\mathcal{C}$ in the following way. Each transition label consists of operands and variables in $\mathcal{A}$ or $\mathcal{\neg A}$, which is shared across the system and specification. Based on this observation, we join graphs $\mathcal{G}$ and $\mathcal{T}$ by adding a link between the corresponding nodes $V_e \in \mathcal{G}$ and $V_\mathcal{T} \in \mathcal{T}$ if they contain the same variable/operand that belongs to $\mathcal{A}$ or $\mathcal{\tilde A}$/$\mathcal{\neg A}$. Figure \ref{fig:Uni} shows the result of such graph union between $\mathcal{G}$ and $\mathcal{T}$. Here, there is an edge between $a$ and $E_{1}$ as $a$ is contained in $E_{1}$. Similarly, there is one edge between $!b$ and $E_{1}$, and the other between $!b$ and $E_{3}$ as $b$ is in $E_{1}$ and $\neg b$ is in $E_{3}$.

Formally, the unified framework is a joint graph $\mathcal{C} = (V_{\mathcal{C}},E_{\mathcal{C}})$ represented as the union of $\mathcal{G}$ and $\mathcal{T}$, where $V_{\mathcal{C}} = V_{\mathcal{G}} \cup V_{\mathcal{T}}$, and $E_{\mathcal{C}} \subseteq V_{\mathcal{C}} \times V_{\mathcal{C}}$ such that, there is an edge between every leaf node $l \in V_{\mathcal{T}}$, and nodes $E \in V_{e}$ such that $l$ contains a variable $a \in \mathcal{A}$ or $!a \in \mathcal{\tilde A}$, and $E$ also contains the same/equivalent variable $a \in \mathcal{A}$ or $\neg a \in \mathcal{\neg A}$.

\vspace{-3mm}
\subsection{Node Encoding and Learning Unified Graph\label{sec:encoding}} 
Each node $v \in \mathcal{C}$ consisting of operands/variables is represented by a vector as follows
\vspace{-2mm}
\[
      \{\substack{\RN{1} \\ \underbrace{[\_]}_{1/0}}
      \substack{\RN{2} \\ \underbrace{[\_,\_,\_,....,\_]}_{\forall a \in \mathcal{A}}}
      \substack{\RN{3} \\ \underbrace{[\_,\_,\_,....,\_]}_{!a/\neg a, \forall a \in \mathcal{A}}}
      \substack{\RN{4} \\ \underbrace{[\_,\_,\_,....,\_]}_{ \forall o \in \{\mathcal{O} / !\}}}
      \substack{\RN{5} \\ \underbrace{[\_,\_]}_{q \in Q}} 
      \substack{\RN{6} \\ \underbrace{[\_,\_]}_{q \in Q} \}}
\]

Part I of 1 bit is reserved for the special variable \textbf{true}($1$) or \textbf{false}($N$). Part II encodes $\forall a \in \mathcal{A}$, with size of $|\mathcal{A}|$. Part III of size $|\mathcal{A}|$ encodes variables/operands in either $\mathcal{\neg A}$ or $\mathcal{\tilde A}$, as both of them are semantically equivalent. Part IV corresponds to operators in $\mathcal{O}$ except $!$, with the size of $|\mathcal{O}|-1$. Part V represents the type of state $q$ of $B$ in 2 bits, where the first bit is true if $q$ is an initial state, and the second bit is true if $q$ is a final state. Part VI represents the source and destination vertex of $B$, as $\mathcal{G}$ is undirected in nature. Parts I through V use one-hot encoding for indication, and part VI uses the source and destination vertex numbers. Part VI remains 0 other than vertices $v \in V_{e}$, as they correspond to the transitions of $B$ which have a source and destination. The difference between the components in parts I through V is thus captured by their positions in the vector.

GNNs as a powerful tool can capture both structural information and node features for graph-structured data through propagation and aggregation of information through message passing, which is ideal for exploiting the structural correspondence between the $B$ and $\phi$, in addition to the semantics of transitions. Graph Isomorphism Network (GIN, \cite{xu2018powerful}), one of the most expressive GRL models, is employed to learn the representation of the unified framework $\mathcal{C}$. The key intuition here is to jointly learn the representation $\mathcal{G}$ and $\mathcal{T}$. Since two structurally similar $B$'s (or $\phi$'s) can represent different behaviors depending on the contents of the transition, both structure and labels that describe the semantics of $B$ (or $\phi$) are equally important for LTL model checking. Modeling the system or the specification separately would lose the crucial connection between them, which is  the essential component in traditional model checkers formed as graph products. Hence, we deploy GIN to capture both structure and semantics of $B$ and $\phi$ jointly in the representation learnt, and the significance of the joint representation for $\mathcal{C}$ is further solidified in Sections \ref{sec:perf} and \ref{sec:case}.
\vspace{-2mm}

\section{Evaluation \label{sec:exp}}
\subsection{Architecture and Hyperparameters}
The architecture of \proj (Figure \ref{fig:OCTAL}, Appendix \ref{appx:exp}) comprises a three-layer GNN. Mean pooling is used to aggregate the learned node embeddings. A dropout rate \cite{gal2016dropout} of 0.1 is used, along with 1D batch normalization \cite{batchnorm} in every convolution layer of GNN. ReLU \cite{RelU18} is used as the non-linear activation between GNN and MLP layers. Every node in $\mathcal{C}$ has an initial embedding of length 66. The GNN framework produces an embedding for $\mathcal{C}$ of dimension 128 after mean pooling. MLPs take this hidden graph representation as the input and produce a `0/1' result as the final prediction.

\subsection{Datasets \label{sec:data}}
We present four datasets namely, \texttt{SynthGen}, \texttt{SynthSpec}, \texttt{RERSGen} and \texttt{RERSSpec}. The datasets constitute of datapoints of the form $(B,\phi,l)$, where $B$ is the system, $\phi$ is the specification and $l$ is the label. $l$ is `1' if $B$ satisfies $\phi$ and `0' otherwise. \texttt{SynthGen} and \texttt{RERSGen} correspond to the general model checking scenario, while \texttt{SynthSpec} and \texttt{RERSSpec} correspond to the special model checking scenario. The datasets \texttt{SynthGen} (\texttt{RERSGen}) and \texttt{SynthSpec} (\texttt{RERSSpec}) share the same $B$ and $\phi$, differing in $l$. Hence, we refer to the distribution of \texttt{Synth(Gen|Spec)} as \texttt{Synth}, and \texttt{RERS(Gen|Spec)} as \texttt{RERS}. \texttt{Synth} is constructed synthetically and \texttt{RERS} is adopted from the specifications of the RERS model checking competition 2019. The motivation behind constructing the mentioned datasets are to test the accuracy, speedup and generalizability of \proj when it comes to complex, lengthy and varied length specifications and/or systems in the same dataset, which previous ML based works failed to solve. The purpose of specifically designing the synthetic dataset was to incorporate a diverse set of specifications, which is different from RERS as the latter caters to traditional model checking competitions where MCs can't handle the length and complexity of specifications/systems beyond a threshold. The statistics and construction of the datasets are summarized in Table \ref{tab:data}, and detailed in Appendix \ref{appx:data}.

\begin{table}[tp]
\centering
\caption{Statistics of \texttt{Synth} and \texttt{RERS}.}
\vspace{-3mm}
\label{tab:data}
\centering
\resizebox{0.7\columnwidth}{!}{
  \begin{tabular}{lccc}
    \toprule
    \textbf{Dataset}&\textbf{Len\_LTL} &\textbf{\#State}&\textbf{\#Transition} \\
    \midrule
    \texttt{Synth} &  [1 - 80]  & [1 - 95] & [1 - 1,711]  \\
    \texttt{RERS} & [3 - 39] & [1 - 21] & [3 - 157] \\
  \bottomrule 
\end{tabular}}
\vspace{-6mm}
\end{table}

\subsection{Experimental Settings \label{sec:settings}}
\paragraph{Training}
\model is trained with an 80-20 split between training and validation sets, which contain \emph{equal} number of positive and negative samples for classification and are randomly shuffled. We use Adam \cite{AdamOptimizer} with  initial learning rate \texttt{lr=1e-5}, and the Binary Cross Entropy \cite{ruby2020binary} as the loss function for all  experiments. Early stopping is adopted when the highest accuracy on validation no longer increases for five consecutive checkpoints. All experiments are run 5 times independently, and the average performance and standard deviations for accuracy, precision, and recall are reported.
\vspace{-2mm}
\begin{figure*}
\subfigure[$B$ implies $\phi$ \label{fig:imply}]{
\begin{minipage}{0.08\textwidth}
\centering    
\includegraphics[width=2.7cm]{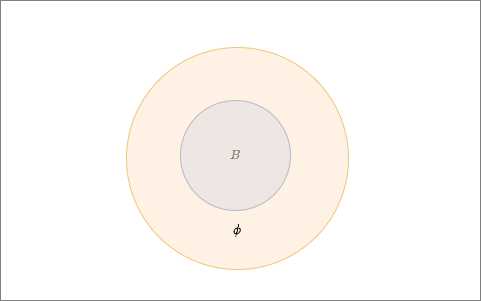}
\end{minipage}
}
\hspace{12mm}
\subfigure[$B$ and $\phi$ are equivalent \label{fig:equiv}]{
\begin{minipage}{0.14\textwidth}
\centering
\includegraphics[width=3.5cm]{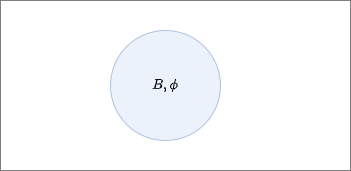}
\end{minipage}
}
\hspace{10mm}
\subfigure[$B$ and $\phi$ are not disjoint \label{fig:intersection}]{
\begin{minipage}{0.14\textwidth}
\centering
\includegraphics[width=3.5cm]{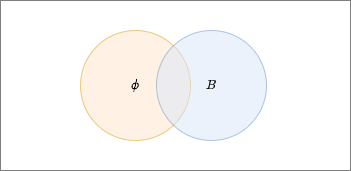}
\end{minipage}
}
\hspace{10mm}
\subfigure[$B$ and $\phi$ are disjoint \label{fig:disjoint}]{
\begin{minipage}{0.14\textwidth}
\centering
\includegraphics[width=3.5cm]{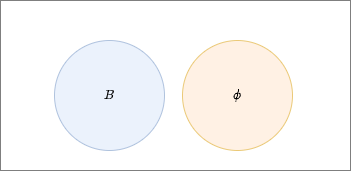}
\end{minipage}
}
\hspace{10mm}
\subfigure[$\phi$ implies $B$\label{fig:noImply}]{
\begin{minipage}{0.1\textwidth}
\centering
\includegraphics[width=2.7cm]{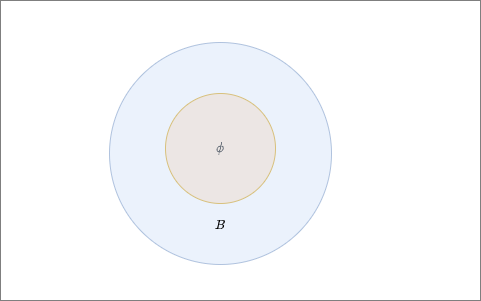}
\end{minipage}
}
\vspace{-7mm}
\caption{(a-b) represent the two scenarios where $B$ satisfies $\phi$. (c-d) represent the scenarios where $B$ does not satisfy $\phi$.}
\vspace{-4mm}
\end{figure*}
\vspace{-2mm}
\paragraph{Baselines} Two classes of methods are selected to compare for LTL model checking:

\textit{Traditional Model Checkers} LTL3BA, Spin \cite{spinMC} and Spot are the SOTA tools that perform traditional symbolic model checking. LTL3BA is the fastest tool among them to compute $B$ for a $\phi$. We select LTL3BA and Spot as the baselines for the speed test due to their superior performance, and run it for every $(B, \phi)$. Since $B$ corresponds to a specification $\phi'$ as per our dataset description, to model check $(B, \phi)$ using LTL3BA, we provide the LTL formula $(\phi' \: \& \: !\phi)$ as input to LTL3BA and verify whether the automaton generated is empty. To model check $(B, \phi)$ using Spot, we check whether $\phi'$ implies $\phi$ through its command line interface. 

\textit{Learning-based Models} Three of the four neural networks take the unified framework $\mathcal{C}$ as input, and outputs `0/1':

\textbf{MLP} A multilayer perceptron (MLP, \cite{MLPCite}) classifier directly uses node features as input without utilizing the unified framework $\mathcal{C}$.

\textbf{LinkPredictor} Graph Convolutional Network (GCN, \cite{kipf2016semi}) is used to learn representations of $\mathcal{G}$ and $\mathcal{T}$ \emph{separately}. The obtained node embeddings are then concatenated and fed into linear layers for classification.

\textbf{\proj} A GRL framework that reduces the model checking to a graph classification task by \emph{jointly} learning the representation of $\mathcal{G}$ and $\mathcal{T}$ through the unified framework $\mathcal{C}$.

\begin{table}
\caption{Classification Accuracy, Precision and Recall for general model checking scenario. \texttt{SynthGen} represents \texttt{SynthGen} used for train and test. \texttt{RERSGen} represents \texttt{SynthGen} used for train and \texttt{RERS} for test.}
\label{tab:acc}
\vspace{-2mm}
\centering
\renewcommand{\arraystretch}{1.15} 
\resizebox{1\columnwidth}{!}{
\begin{tabular}{ll|cc|cc|cc}
\toprule
    & \multirow{2}{*}{\textbf{Models}} & \multicolumn{2}{c|}{\textbf{Accuracy}} & \multicolumn{2}{c|}{\textbf{Precision}} & \multicolumn{2}{c}{\textbf{Recall}} \\ \cmidrule(r{0.5em}){3-8}
    &  & \texttt{SynthGen}  & \texttt{RERSGen}  & \texttt{SynthGen}  & \texttt{RERSGen} & \texttt{SynthGen}  & \texttt{RERSGen} \\
\midrule
     & \textbf{MLP} &  46.44±0.86 & 51.73±1.38 & 46.83±0.64 & 51.72±1.46 & 52.19±3.66 & 48.01±9.36 \\
    & \textbf{LinkPredictor} & 60.76±0.81 & 67.93±1.18 & 61.86±0.92 & 66.66±1.39 & 56.12±0.81 & 71.83±1.17 \\
    & \textbf{OCTAL(GCN)} & 76.76±0.95 & 88.23±0.75 & 84.77±1.20 & 88.97±1.11 & 65.26±2.27 & 87.32±2.21 \\
    & \textbf{OCTAL(GIN)} & \textbf{77.96±1.71} & \textbf{89.48±0.61} & \textbf{85.37±1.11} & \textbf{89.63±2.17} & \textbf{67.51±3.99} & \textbf{89.37±1.73} \\
\bottomrule
\end{tabular}}
\vspace{-5mm}
\end{table}

\vspace{-2mm}
\subsection{Performance Analysis \label{sec:perf}}
Table \ref{tab:acc} shows the performance of different methods for LTL MC as a classification task. \proj(GIN) consistently outperforms all the other baselines, and achieves $\sim$ 90\% accuracy on \texttt{RERSGen} and $\sim$ 78\% accuracy on \texttt{SynthGen}. \proj(GCN) reports high accuracy but is slightly behind \proj(GIN) on average due to GCN's limited expressiveness compared to GIN. In general, message passing frameworks outperform MLP and LinkPredictor, which either do not take graph structures into account or model $B$ and $\phi$ separately. 

To better evaluate the models, we consider the metrics of precision and recall. From Table \ref{tab:acc}, we see that for \texttt{RERSGen}, all three metrics have similar values. However, for \texttt{SynthGen}, we observe a stark difference between precision and recall. The precision reported is $\sim85\%$, which is $\sim8\%$ higher than the accuracy, and the recall reported is $\sim67\%$, which is $\sim11\%$ lower than the accuracy. The results imply great stability and generalization of \model for \texttt{RERSGen}. A detailed analysis of precision and recall results is presented in Appendix \ref{appx:fanalysis}.

\begin{table}
\caption{Speedup for general LTL model checking. \textbf{LT3BA (Spot)} (I) represents inference-only speedup. \textbf{LT3BA (Spot)} (O) represents the sum of inference and graph construction overhead speedup.}
\label{tab:speedupgen}
\vspace{-3mm}
\centering
\resizebox{0.8\columnwidth}{!}{
\begin{tabular}{l|cc|cc}
\toprule
    \textbf{Dataset} &  \textbf{LTL3BA(I)} & \textbf{LTL3BA(O)} & \textbf{Spot(I)} & \textbf{Spot(O)} \\
    \midrule
    \texttt{SynthGen} & $351\times$ & $11\times$ & $52\times$ & $1.7\times$  \\
    \texttt{RERSGen} & $54\times$ & $2\times$ & $30.5\times$ & $1.6\times$ \\
\bottomrule
\end{tabular}}
\vspace{-6mm}
\end{table}

\subsection{Complexity Analysis \label{sec:complex}}
Table \ref{tab:speedupgen} shows the speedup of \proj with respect to LTL3BA and Spot, for inference only, and inference with graph construction overhead. Compared to traditional MCs, with preprocessing overhead considered, NN-based models are still $\sim11\times$ faster than LTL3BA and $\sim1.7\times$ faster than Spot for \texttt{SynthGen}. NN-based models are $\sim2\times$ faster than LTL3BA and Spot for \texttt{RERSGen}. It is worth noting that, in terms of inference alone, \proj is $\sim351\times$, $\sim54\times$ faster than LTL3BA, and $\sim52\times$, $\sim31\times$ faster than Spot on both datasets. Hence, we infer that \proj can outperform the SOTA traditional model checkers with respect to speed, along with consistent accuracy across different datasets. We also conclude that \proj tends to have an increase in speedups over traditional MCs when it comes to more complex and lengthy specifications/systems, as from the distribution, we see that \texttt{Synth} subsumes \texttt{RERS}. Note that, as a proof of concept, the graph preprocessing time presented above is not extensively optimized in terms of speed. We aim to provide a parallel graph construction algorithm in the future that would significantly reduce the preprocessing overhead.

Traditional model checkers map $\neg \phi$ to $B_{\neg \phi}$, compute the product $B' = B_{\neg \phi} \times B$, and then check emptiness of $B'$. The use of the union operation pairing with GRL framework enables \proj to avoid the non-polynomial complexity of graph product. Accordingly, the complexity of our proposed method is reduced to polynomial in the size of $|B + \phi|$. 
\newline

\begin{table}
\caption{Classification Accuracy, Precision and Recall for special model checking scenario for \texttt{SynthSpec} and \texttt{RERSSpec}}.
\label{tab:specialMCResult}
\vspace{-3mm}
\centering
\renewcommand{\arraystretch}{1.15} 
\resizebox{1\columnwidth}{!}{
\begin{tabular}{ll|cc|cc|cc}
\toprule
    & \multirow{2}{*}{\textbf{Models}} & \multicolumn{2}{c|}{\textbf{Accuracy}} & \multicolumn{2}{c|}{\textbf{Precision}} & \multicolumn{2}{c}{\textbf{Recall}} \\ \cmidrule(r{0.5em}){3-8}
    & & \texttt{SynthSpec} & \texttt{RERSSpec}  & \texttt{SynthSpec}  & \texttt{RERSSpec} & \texttt{SynthSpec}  & \texttt{RERSSpec} \\
\midrule
     & \textbf{MLP} &  48.90±0.80 & 59.53±1.66 & 48.90±0.75 & 59.07±1.10 & 45.39±2.86 & 61.96±5.67 \\
    & \textbf{LinkPredictor} & 73.13±1.11 & 73.54±1.98 & 72.39±0.98 & 70.02±1.98 & 74.87±2.80 & 82.41±2.84 \\
    & \textbf{OCTAL(GCN)} & 95.18±0.47 & 95.45±0.72 & \textbf{95.32±0.71} & 91.82±1.02 & 95.03±0.76 & \textbf{99.81±0.29} \\
    & \textbf{OCTAL(GIN)} & \textbf{95.37±0.69} & \textbf{96.19±0.62} & 94.57±1.39 & \textbf{95.52±0.69} & \textbf{96.30±0.68} & 96.94±1.91 \\
\bottomrule
\end{tabular}}
\end{table}
\vspace{-6mm}
\begin{table}
\caption{Speedups for special LTL model checking.}
\label{tab:speedupspcl}
\vspace{-4mm}
\centering
\resizebox{0.8\columnwidth}{!}{
\begin{tabular}{l|cc|cc}
\toprule
    \textbf{Dataset} &  \textbf{LTL3BA(I)} & \textbf{LTL3BA(O)} & \textbf{Spot(I)} & \textbf{Spot(O)} \\
    \midrule
    \texttt{SynthSpec} & $282\times$ & $9.3\times$ & $49\times$ & $1.6\times$  \\
    \texttt{RERSSpec} & $37.3\times$ & $2\times$ & $30.5\times$ & $1.6\times$ \\
\bottomrule
\end{tabular}}
\vspace{-5mm}
\end{table}

\subsection{Case Study: On the equivalence of $B$ and $\phi$ \label{sec:case}}
We evaluate \proj on a special case of model checking, where $B$ accepts $\phi$ iff they are equivalent (Figure \ref{fig:equiv}). The goal of this setting is to evaluate \proj on system equivalence which is indeed a hard problem. Tables \ref{tab:specialMCResult} and \ref{tab:speedupspcl} show the accuracy and speedup results. Here too, the message passing networks outperform the MLP and LinkPredictor, signifying the importance of learning the union of $\mathcal{G}$ and $\mathcal{T}$. The results for equivalence checking for \proj are very impressive. \proj consistently reports accuracy $\sim95\%$ across \texttt{SynthSpec} and \texttt{RERSSpec} which is significantly higher than the general case. Further experiments on runtimes obtained for the datasets corresponding to general and special model checking is presented in Appendix \ref{appx:fanalysis}.
\vspace{-2mm}
\section{Related Work \label{sec:Related}}
To the best of our knowledge, this is the first work that applies graph representation learning to solve LTL model checking. Previously, the most relevant work to us was using GNNs to solve SAT for boolean satisfiability \cite{neuroSAT}, and automated proof search \cite{HOLGraph} in the higher-order logic space. Learning-based approaches have been used to select the most suitable model checker \cite{MUXTulsianKKLN14}. \cite{IntegrateBorgesGL10} attempts to learn how to reshape a system \cite{cimatti1999nusmv} to satisfy the property in temporal logic. \cite{TemporalTransformer} trains a transformer to predict a satisfiable trace for an LTL formula. \cite{ZhuMC} proposes ML-based model checking, with the system being Kripke structures and specification being LTL formulae, both serving as input features to supervised ML algorithms. Their framework is shown to perform well with formulae of the same lengths, otherwise not. \cite{BehjatiRI} proposes a reinforcement learning based approach for on-the-fly LTL model checking, which is designed to look for invalid runs or counterexamples by awarding heuristics with an agent. Their approach performs faster than the classical model checkers and can verify systems with large state spaces, but the state space that an agent can reach is still bounded.
\vspace{-2mm}
\section{Conclusions and Future Work \label{sec:Conclusion}}
\proj is a novel graph representation learning based framework for LTL model checking. It can be extremely useful for the first line of the software development cycle, as it offers reasonable accuracy and robustness for early and quick verification, compared to time-consuming unit tests and other efforts in ensuring the correctness of a given system. \model is not intended to replace traditional model checkers, it rather makes model checking affordable and scalable for scenarios where traditional model checkers are infeasible.  It can also be enhanced with a guarantee provided by applying traditional model checkers to limited candidates filtered by \model.  In future, we propose to improve \proj on the false negatives for implication cases and extend \proj to support the generation of a counter-example trace for the `no' answers. Since the counter-example generation is a relatively easier problem, tentatively, the user can invoke a traditional model checker to obtain it.
\vspace{-1mm}
\section{Acknowledgements \label{sec:Ack}}
The authors sincerely thank Professor Tiark Rompf and Dr. Susheel Suresh for their valuable feedback throughout the project.

\bibliographystyle{ACM-Reference-Format}
\bibliography{ref}

\appendix
\newpage
\section{Traditional Model Checking and Temporal Operators \label{appx:symb}}

\subsection{B{\"u}chi Automata}
In automata theory, a B{\"u}chi automaton (BA) is a system that either accepts or rejects inputs of infinite length. The automaton is represented by a set of states (one initial, some final, and others neither initial nor final), a transition relation, which determines which state should the present state move to, depending on the alphabets that hold in the transition. The system accepts an input if and only if it visits at least one accepting state infinitely often for the input. A B{\"u}chi automaton can be deterministic or non-deterministic. We deal with non-deterministic B{\"u}chi automata systems in this paper, as they are strictly more powerful than deterministic B{\"u}chi automata systems.

A non-deterministic B{\"u}chi automaton $B$ is formally defined as the tuple $(Q, \sum, \Delta, q_{0}, \mathbf{q}_{f})$, where $Q$ is the set of all states of $B$, and is finite; $\sum$ is the finite set of alphabets; $\Delta: Q \times 2^{\sum} \rightarrow 2^Q$ is the transition relation that can map a state to a set of states on the same input set; $q_{0} \in Q$ represents the initial state; $\mathbf{q}_{f} \subset Q$ is the set of final states.

\subsection{Linear Temporal Logic}
Linear Temporal Logic (LTL), is a type of temporal logic that models properties with respect to time. An LTL formula is constructed from a finite set of atomic propositions, logical operators not (!), and (\&), and or ($|$), true (1), false ($N$), and temporal operators. Additional logical operators such as implies, equivalence, etc. that can be replaced by the combination of basic logical operators $(!, \&, |)$. For example, $a \rightarrow b$ is expressed as $!a \: | \: b$.

\subsection{LTL Model Checking}
Given a B{\"u}chi automaton $B$ (system), and an LTL formula $\phi$ (specification), the model checking problem decides whether $B$ satisfies  $\phi$. Traditionally, the problem is solved by computing the B{\"u}chi automaton for the negation of the specification $\phi$ as $B_{\neg \phi}$, followed by the product automaton $B' = B \times B_{\neg \phi}$. The problem then reduces to checking the language emptiness of $B'$. The language accepted by $B'$ is said to be empty if and only if $B'$ rejects all inputs. Construction of $B$ is linear in the size of its state space, while $B_{\neg \phi}$ is exponential in the size of $\neg \phi$. The product construction would also lead to an automaton of size $|B| \times 2^{|\neg \phi|}$, which can blow up even $\phi$ is linear in the size of $B$, leading to the state space explosion problem. Figure \ref{fig:BA} represents the system $B$ with $\phi$: `$a \: U \: !b$',
and the cross product $B'=B\times B_{\neg \phi}$ is given in Figure \ref{FigurelabelBACrossProd}. $B'$ does not accept anything as there is no feasible path from the initial state ($q'_{0}q_{0}$) to either of the final states ($q'_{0}q_{f}$, $q'_{f}q_{f}$). Here, both $B$ and $B_{\neg \phi}$ are three times smaller than $B'$ in terms of the number of states, and 6 times smaller regarding the number of transitions. As observed in Figure \ref{FigurelabelBACrossProd}, it can be concluded that even for moderately complex specifications, the product can still result in an exponential state space, which would severely hinder the performance of traditional model checkers.

\begin{table}
\centering
\caption{Temporal Operators in Linear Temporal Logic}
\label{tab:symbol}
\centering
\resizebox{1.0\columnwidth}{!}{
  \begin{tabular}{lcccccccc}
    \toprule
    \textbf{Symbol}&G&F&R&W&M&X&U\\
    \midrule
    \textbf{Meaning} & globally & finally & release & weak until & strong release & next & until \\
  \bottomrule 
\end{tabular}}
\vspace{-3mm}
\end{table}

The meaning of temporal operators supported by $\phi$ is presented in Table \ref{tab:symbol}. 

\section{Experimental Settings and Datasets \label{appx:exp}}
\subsection{Environment} 
Experiments were performed on a cluster with four Intel 24-Core Gold 6248R CPUs, 1TB DRAM, and eight NVIDIA QUADRO RTX 6000 (24GB) GPUs. 

\begin{figure}[htp]
    \centering
    \includegraphics[width=0.48\textwidth]{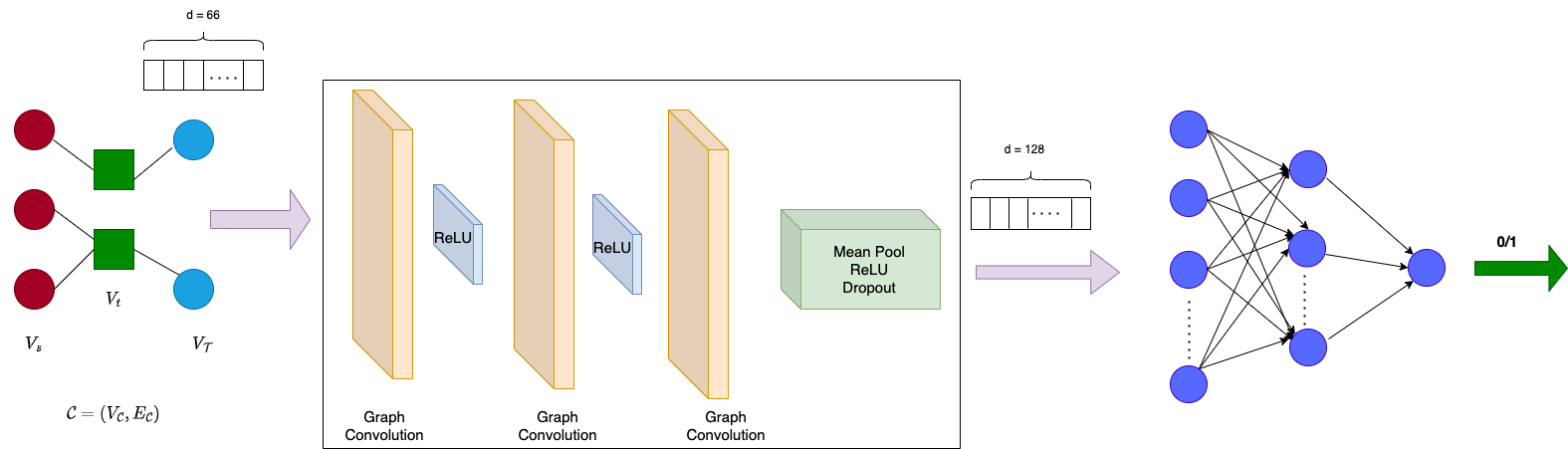}
    \caption{Overview of \proj Architecture.}
    \label{fig:OCTAL}
\vspace{-3mm}
\end{figure}

\subsection{Implementation Details}
The code base is implemented on PyTorch 1.8.0 and pytorch-geometric \cite{pytorchGeometric}. The source files are attached along with the submitted supplementary.

\subsection{Datasets Statistics of \texttt{Synth} and \texttt{RERS} \label{appx:data}}
 The specifications for \texttt{Synth} are synthetically generated through Spot \cite{spotpaper}, while \texttt{RERS} is obtained from the annual competition RERS19 \cite{rers2019}. The data generated from Spot and provided by RERS19 are specifications, i.e., a set of LTL formulae $\phi$'s. To generate the corresponding automaton $B$ for each $\phi$, the tool LTL3BA \cite{ltl3bapaper} is used, due to its superior speed \cite{ltl3baspeed}. \texttt{Synth} consists of specifications, all of which can be solved by LTL3BA within the time limit (2 mins). \texttt{Synth} aims to test the checking speed of models w.r.t LTL3BA, along with generalization ability for varied length of specifications. \texttt{RERS} is a standard benchmark for sequential LTL from the RERS19 competition. The datasets are catered corresponding to the two scenarios we evaluate \proj on. The details of the construction of yes/no pairs for \texttt{Synth} and \texttt{RERS} corresponding to the two applications are discussed in detail in the following sections.

 \subsubsection{Datasets \texttt{SynthGen} and \texttt{RERSGen}}
These datasets correspond to the general model checking scenario, where the yes instances correspond to the system being equivalent to the specification, and the system strictly implying the specification (Figures \ref{fig:equiv}, \ref{fig:imply}). Hence, every specification $\phi$ is paired with systems $B_{1}$, $B_{2}$, $B_{3}$ and $B_{4}$, where $B_{1}$ is equivalent to $\phi$, $B_{2}$ implies but is not equivalent to $\phi$, $B_{3}$ and $B_{4}$ do not imply $\phi$ (Figures \ref{fig:intersection}, \ref{fig:disjoint}, \ref{fig:noImply}). Hence, $(B_{1},\phi)$ and $(B_{2},\phi)$ have label 1, whereas $(B_{3},\phi)$ and $(B_{4},\phi)$ have label 0. For \texttt{SynthGen}, $\phi$ and $B$ correspond to \texttt{Synth}, and for \texttt{RERSGen} $\phi$ and $B$ correspond to \texttt{RERS}.

\subsubsection{Datasets \texttt{SynthSpec} and \texttt{RERSSpec}}
These datasets correspond to the special model checking scenario, where we evaluate \proj on a strict subset of model checking, where the system satisfies the specification, iff they are equivalent. Hence, every $\phi$ is paired with systems $B_{1}$ and $B_{2}$, where $B_{1}$ is equivalent to $\phi$ and $B_{2}$ is not. Hence, $(B_{1},\phi)$ has label 1 and $(B_{2},\phi)$ has label 0. We consider an equal distribution of strict implication (\ref{fig:imply}) and other and non-implication (\ref{fig:intersection}, \ref{fig:disjoint}, \ref{fig:noImply}) tuples for the 0 cases. For \texttt{SynthSpec}, $\phi$ and $B$ correspond to \texttt{Synth}, and for \texttt{RERSSpec} $\phi$ and $B$ correspond to \texttt{RERS}.

\subsubsection{\texttt{Synth} generation details}
\begin{figure*}
\subfigure[Length distribution \label{fig:lenSynth}]{
\begin{minipage}{0.45\textwidth}
\centering    
\includegraphics[height=4cm]{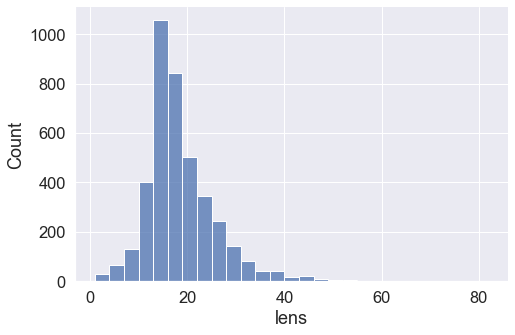}
\end{minipage}
}
\subfigure[Length v.s. State distribution \label{fig:stateSynth}]{
\begin{minipage}{0.45\textwidth}
\centering
\includegraphics[height=4.5cm]{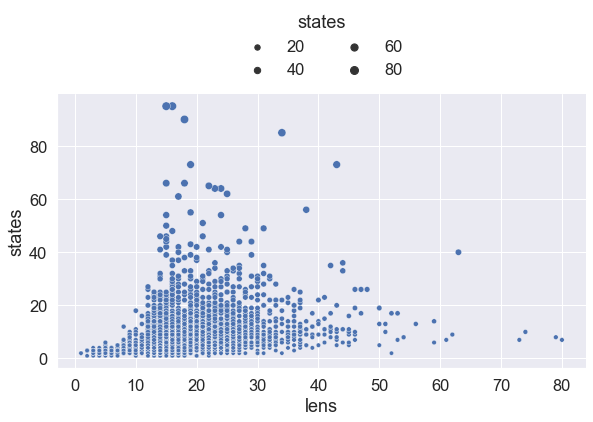}
\end{minipage}
}

\subfigure[Length v.s. Edge distribution \label{fig:edgeSynth}]{
\begin{minipage}{0.45\textwidth}
\centering    
\includegraphics[height=4cm]{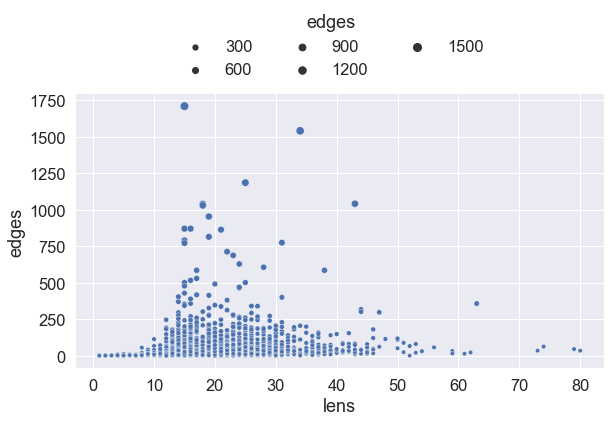}
\end{minipage}
}
\subfigure[Edge v.s. State distribution \label{fig:stateEdgeSynth}]{
\begin{minipage}{0.45\textwidth}
\centering
\includegraphics[height=4.5cm]{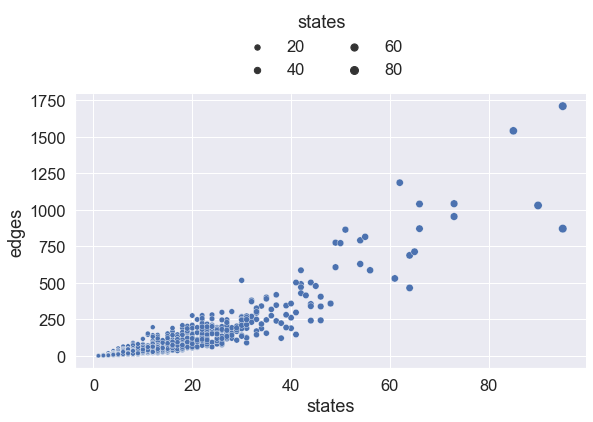}
\end{minipage}
}
\caption{Statistical Summary of \texttt{Synth}.}
\label{fig:scalingSynth}
\end{figure*}

The \texttt{randltl} feature of Spot controls the length of generated LTL formulae, where the default size of expression tree is set to 15. The output formulae are not syntactically the same . The specifications generated through Spot have LTL formulae of lengths (noted as \#Lens) ranging from 1 to 80. The length distribution of the formula for \texttt{Synth} is plotted in Figures \ref{fig:lenSynth}. The number of states and edges are described in Figures \ref{fig:stateSynth}, \ref{fig:edgeSynth}, respectively. By observing those distributions, it can be concluded that the range of length and states of LTL formulae of \texttt{RERS} is subsumed by \texttt{Synth} The corresponding transition range is less than 160 for \texttt{RERS} while 1,711 for \texttt{Synth}.

\begin{figure*}
\subfigure[Length distribution \label{fig:lenRERS}]{
\begin{minipage}{0.45\textwidth}
\centering
\includegraphics[height=4.5cm]{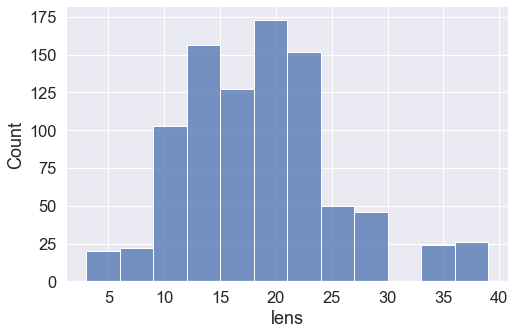}
\end{minipage}
}
\hfill
\subfigure[Length vs State distribution \label{fig:stateRERS}]{
\begin{minipage}{0.45\textwidth}
\centering
\includegraphics[height=4.5cm]{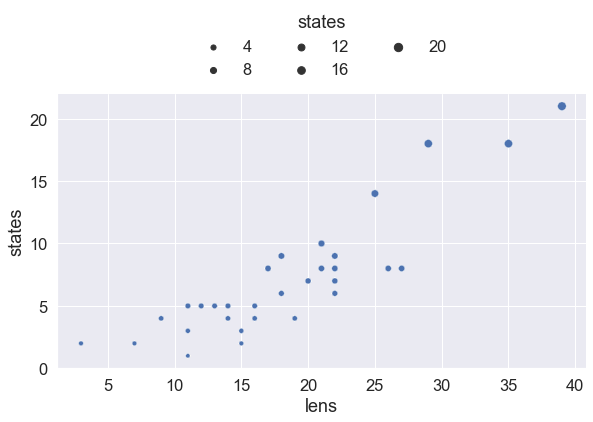}
\end{minipage}
}
\hfill
\subfigure[Length vs Edge distribution \label{fig:edgeRERS}]{
\begin{minipage}{0.45\textwidth}
\centering    
\includegraphics[height=4.5cm]{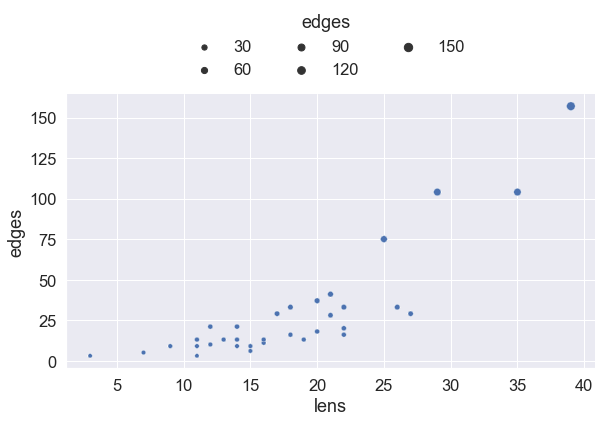}
\end{minipage}
}
\hfill
\subfigure[Edge vs State distribution \label{fig:stateEdgeRERS}]{
\begin{minipage}{0.45\textwidth}
\centering
\includegraphics[height=4.5cm]{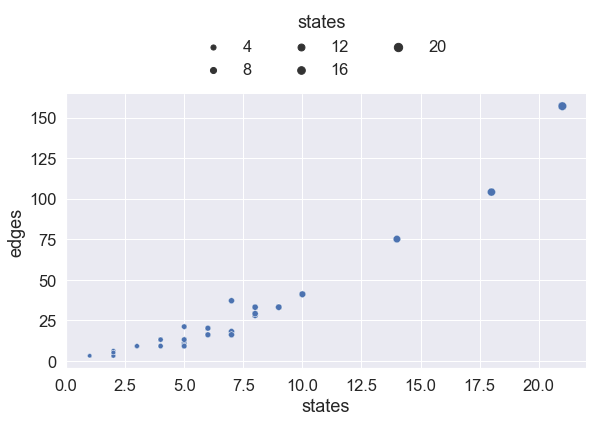}
\end{minipage}
}
\caption{Statistical Summary of  \texttt{RERS}.}
\label{fig:scalingRERS}
\end{figure*}

\subsubsection{\texttt{RERS} generation details}
Rigorous Examination of Reactive Systems (RERS) is an international model checking competition track organized every year. We adopt 900 specifications from the Sequential LTL track, generate the corresponding BAs and construct the dataset for the two cases. The statistical details of \texttt{RERS} is presented in Figure \ref{fig:scalingRERS}.

\section{Further Experiments and Analysis \label{appx:fanalysis}}
\subsection{Generalization on \texttt{SynthGen}}
\begin{table}
\caption{Classification Accuracy, Precision and Recall for general model checking scenario where \texttt{RERSGen} is used for training and \texttt{SynthGen} for inference.}
\label{tab:accApp}
\vspace{-2mm}
\centering
\renewcommand{\arraystretch}{1} 
\resizebox{0.8\columnwidth}{!}{
\begin{tabular}{ll|ccc}
\toprule
    & \textbf{Models} & \texttt{Accuracy} & \texttt{Precision}  & \texttt{Recall} \\
\midrule
    & \textbf{OCTAL} & 75.34±0.99 & 85.97±1.02 & 60.59±3.07 \\
\bottomrule
\end{tabular}}
\vspace{-3mm}
\end{table}
In this experimental setting, we evaluate the performance of out of distribution data. From the distributions of \texttt{Synth} and \texttt{RERS} in section \ref{appx:data}, we observe that \texttt{Synth} strictly subsumes \texttt{RERS}, hence we test how well \proj performs inference on \texttt{SynthGen} when trained with \texttt{RERSGen}. From the results in Table \ref{tab:accApp} we observe that the values of accuracy, precision and recall are similar to the scenario in Table \ref{tab:acc} where both training and inference is done on \texttt{SynthGen}. Hence, we can conclude that it generalizes similarly to out of distribution data, as it does for data points in the same distribution. 

\subsection{Analysis of Precision and Recall}

\begin{figure*}[h]
\subfigure[Incorrect predictions for \texttt{SynthGen} \label{fig:incorrectPred}]{
\begin{minipage}{0.45\textwidth}
\centering    
\includegraphics[scale=0.4]{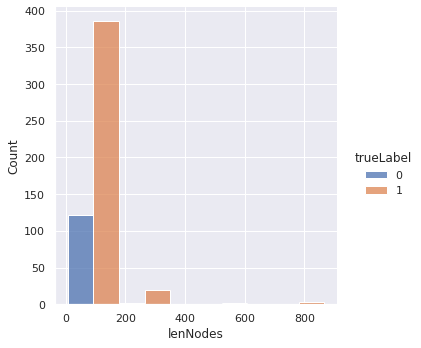}
\end{minipage}
}
\hfill
\subfigure[Correct predictions for \texttt{SynthGen} \label{fig:correctPred}]{
\begin{minipage}{0.45\textwidth}
\centering
\includegraphics[scale=0.4]{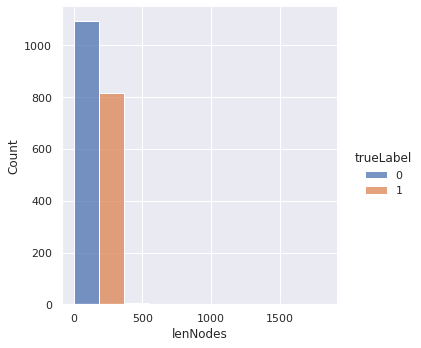}
\end{minipage}
}
\caption{Figures \ref{fig:incorrectPred} and \ref{fig:correctPred} represent the correct and incorrect predictions for \texttt{SynthGen}. The x-axis represents the number of nodes in $\mathcal{G}$ and the y-axis represents the number of miss classified labels. The labels correspond to true labels, hence represent False Positives and False Negatives for \ref{fig:incorrectPred}, and True Positives and True Negatives for \ref{fig:correctPred}.}
\end{figure*}

To better understand the results of \texttt{SynthGen} wrt the difference between Precision and Recall, we plotted the correct (Figure \ref{fig:correctPred}) and incorrect (Figure \ref{fig:incorrectPred}) predictions for the inference of \texttt{SynthGen}. We can observe that the \textit{true} labels have been wrongly classified more than the \textit{false} labels, thus yielding more false negatives which results in a low value of Recall. We further investigated the false negatives and found that all of the miss classified true labels belong to the strict implication case, i.e., the scenario in which the $B$ strictly implies $\phi$ (Figure \ref{fig:imply}). This leads us to conclude that \proj performs and generalizes very well when the distribution of the test set is subsumed by the train set (\texttt{RERSGen}), but has some difficulty in classifying the implications correctly otherwise.


\begin{table}
\label{tab:speedgen}
\caption{Running time of different models for general LTL model checking.}
\label{tab:complexity}
\vspace{-2mm}
\centering
\resizebox{1\columnwidth}{!}{
\begin{tabular}{l|cc|ccc|c}
\toprule
    \textbf{Dataset} &  \textbf{LTL3BA} & \textbf{Spot} & \textbf{MLP} & \textbf{LinkPredictor} & \textbf{\proj} & \textbf{Overhead}\\
    \midrule
    \texttt{SynthGen} & 1154s & 171s & 2.34s & 3.32s & 3.29s & 100.47s \\
    \texttt{RERSGen} & 44s & 34s &0.58s & 0.97s & 0.82s  & 20.67s\\
\bottomrule
\end{tabular}}
\end{table}

\begin{table}
\label{tab:rtspcl}
\caption{Running time of different models for special LTL model checking.}
\label{tab:complexityspcl}
\vspace{-2mm}
\centering
\resizebox{1\columnwidth}{!}{
\begin{tabular}{l|cc|ccc|c}
\toprule
    \textbf{Dataset} &  \textbf{LTL3BA} &  \textbf{Spot} & \textbf{MLP} & \textbf{LinkPredictor} & \textbf{\proj} & \textbf{Overhead}\\
    \midrule
    \texttt{SynthSpec} & 482s & 84s & 1.25s & 1.72s & 1.71s & 50.16s \\
    \texttt{RERSSpec} & 22s & 18s & 0.42s & 0.57s & 0.59s  & 10.79s\\
\bottomrule
\end{tabular}}
\vspace{-3mm}
\end{table}
\end{document}